\documentclass[acus]{JAC2000}

\usepackage{graphicx}

\begin{document}

\title{LATTICE MATCHING WITH A QUADRUPOLE MISSING}
\author{Rainer Wanzenberg, \\
DESY, Notkestr. 85, 
22603 Hamburg, Germany
}
\date{ }

\maketitle

\begin{abstract} 
The lattice for the present design of the TESLA Linear
Collider with integrated X-Ray Laser Facility
is basically a FODO structure with constant
beta-function. There are more than 800 individually
powered superconducting quadrupoles to focus
the beam in the two main linear accelerators (each with a length
of 15 km). For the availability of the beams it is
important that a power supply failure or even a
magnet failure does not cause a significant down time of the
linac. It is demonstrated that the beta-function 
can be matched with a quadrupole "missing". One part
of the main linac (up-to 50 GeV) will be used to accelerate
the High Energy Physics beam and the Free Electron Laser
drive beam with different accelerating gradients.
Therefore the betatron phase advance will be 60$^{\circ}$
or 90$^{\circ}$ depending on the beam energy. It is
shown that even in that case it is possible to match
the lattice simultaneously for both beams with a
quadrupole missing. 
\end{abstract}


\section{Introduction}

The conceptual design of the TESLA linear collider
with integrated x-ray laser facility \cite{CDR}
requires that one of the two 15 km long
super-conducting linear accelerator has to provide
beams for high energy physics (HEP) experiments
and beams for the x-ray free electron laser (FEL)
facility. 
\begin{figure}[h!btp]
\setlength{\unitlength}{1mm}
\centering
\includegraphics*[width=78mm]{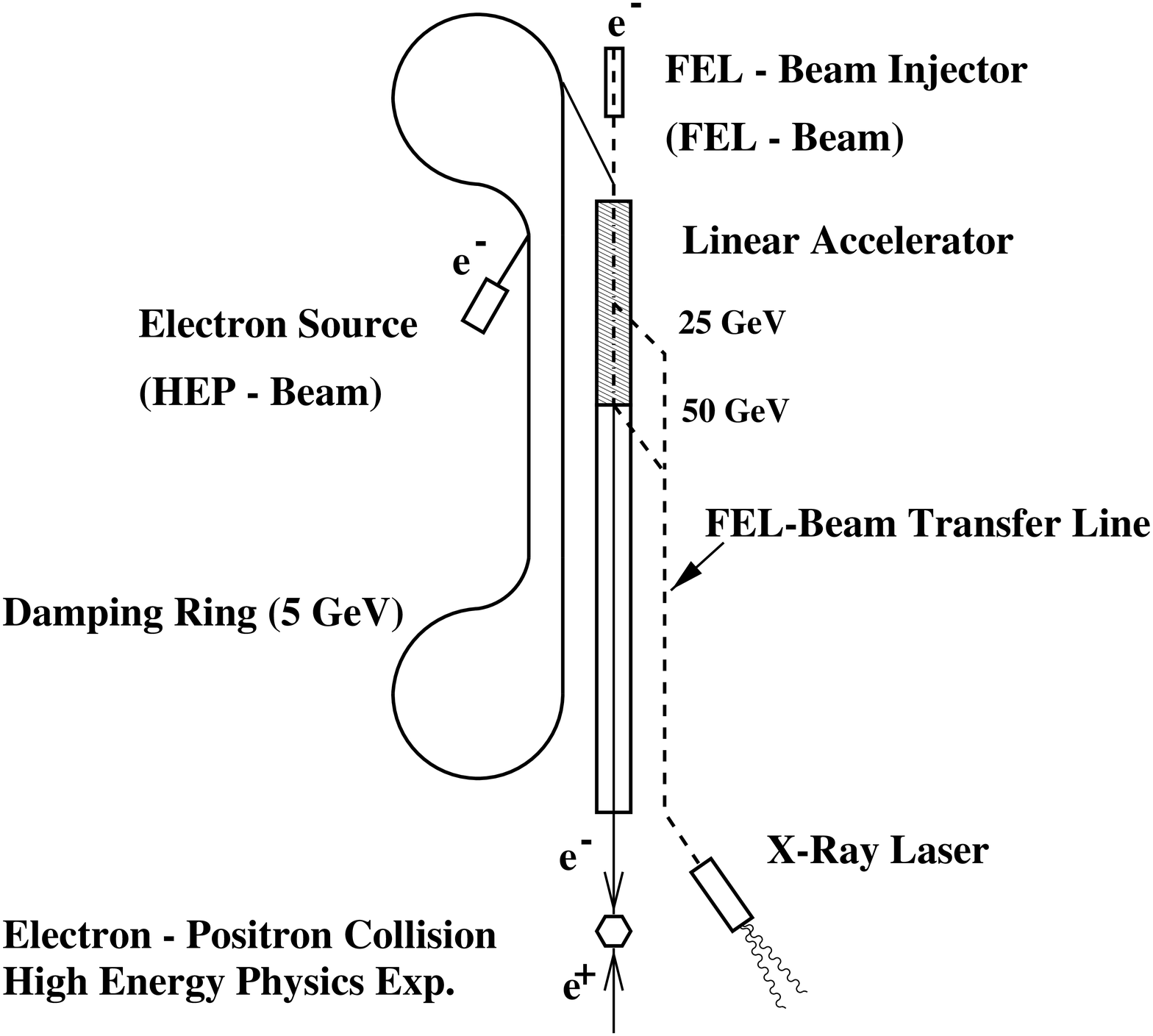}
\caption{The e$^{-}$ linear accelerator of the
TESLA linear collider with integrated x-ray laser
facility. }
\label{overview}
\end{figure}
Fig.~\ref{overview} shows the e$^{-}$
linear accelerator which is used to accelerate the
HEP-beam to 250 GeV and the FEL-beam to energies
up-to 50 GeV. There are 25 GeV and 50 GeV
extraction points in the linac.
 The HEP beam is injected from the
dog-bone damping ring into the main linear 
accelerator while the FEL beam is generated
with an rf-gun and accelerated to 5 GeV in an
injector linac before it is also injected into
the main linear accelerator.

\begin{figure}[h!btp]
\setlength{\unitlength}{1mm}
\centering
\includegraphics*[width=78mm]{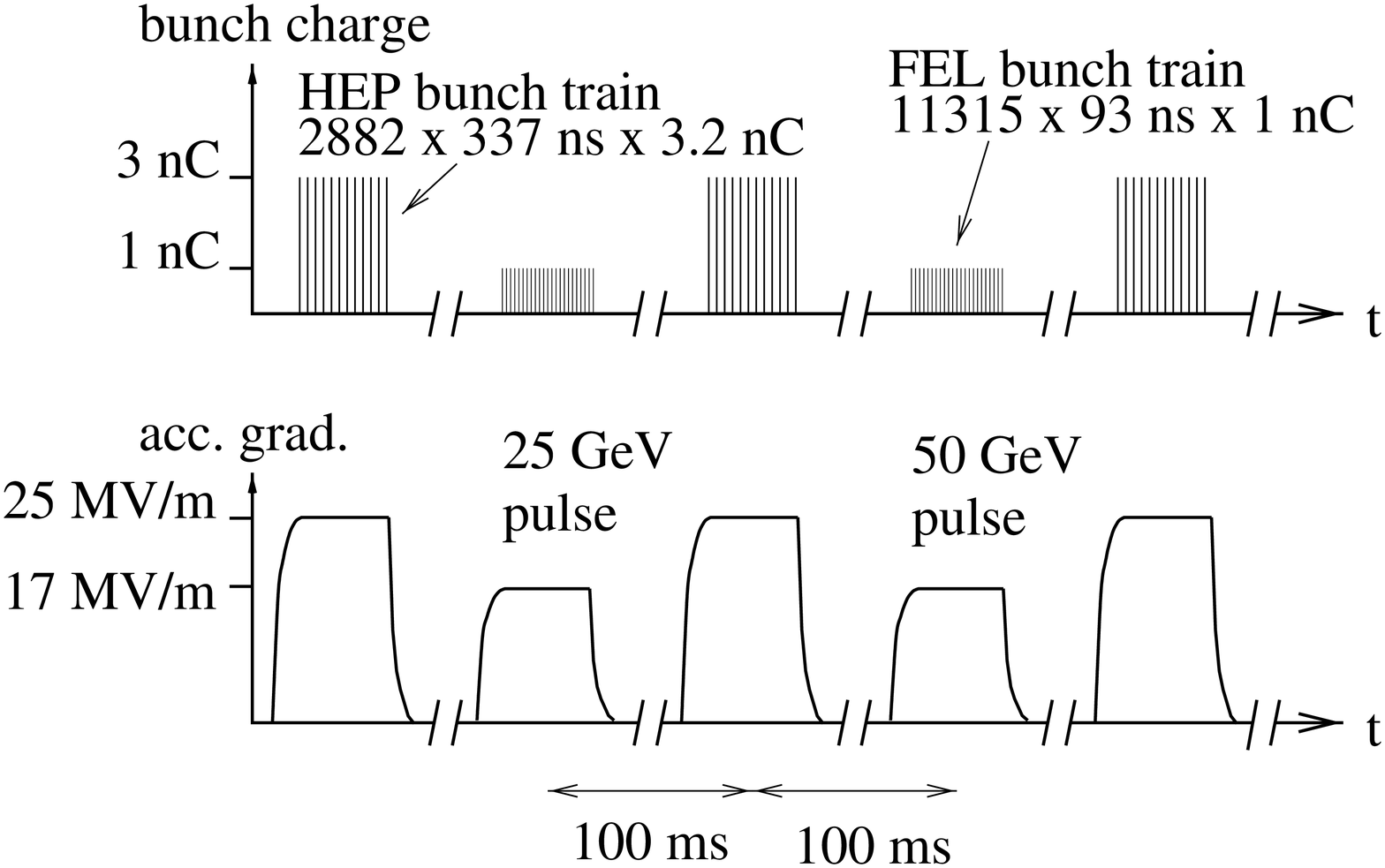}
\caption{HEP and FEL beam pulse structure}
\label{pulse}
\end{figure}

The first part of the e$^{-}$ linear accelerator is
operated at a duty cycle of 10 Hz providing
HEP (2882 bunches) and FEL (11315 bunches) pulses
in a alternating way. The pulse structure is illustrated
in Fig.~\ref{pulse}. The FEL-pulses are accelerated at the
same gradient of 17 MV/m but are ejected at 25 GeV or
50 GeV beam energy to be used for different undulators.

The basic optical building block is a FODO cell with
a length of 65~m for the first half of the linac (up-to 125 GeV)
and a length of 97~m for the second half. Two cryo-modules,
housing 12 (7-cell) cavities each, will fit between two quadrupoles
of the 65~m long FODO cell, while three  cryo-modules will fit between
the quadrupoles of the 97~m long cell.
The quadrupole strength $k=g \, e/p$ ($g$ the gradient of the
magnet, $p$ the momentum of the beam) is different for the
FEL and HEP beams due to the different accelerating gradients:
$ k_{FEL} = k_{HEP} \,\cdot \, 25/17$. A betatron phase advance of
60$^{\circ}$ is chosen for the HEP beam optics. This
corresponds to a phase advance of about 90$^{\circ}$ for
the FEL beam since the focusing is stronger than that of the
HEP beam due to the lower momentum.

For the availability of the beams it is
important that a power supply failure or even a
magnet failure does not cause a significant down time of the
linac. In the next section it will be shown how the lattice
can be matched with a quadrupole "missing" in the case
of the 60$^{\circ}$-FODO lattice. Finally it is demonstrated
that it is possible to match the lattice {\em simultaneously }
for both beams.


\section{60$^{\circ}$-FODO Lattice with a Quadrupole Missing }

\begin{figure}[h!btp]
\setlength{\unitlength}{1mm}
\centering
\includegraphics*[width=80mm]{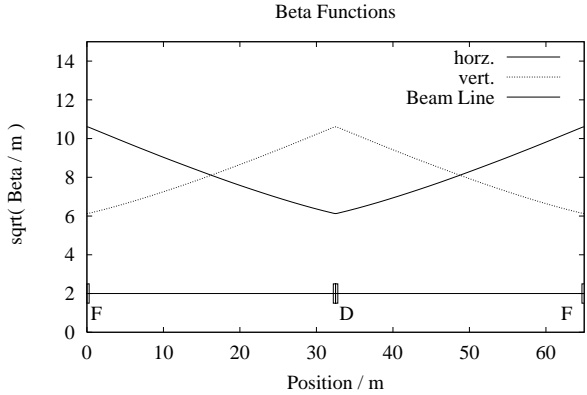}
\vspace*{-5mm}
\caption{Standard FODO cell of the linac (60$^{\circ}$ phase advance).
The square root of the horizontal (solid line) and vertical
(dashed line) beta-function are shown versus the longitudinal position
for the HEP beam.
}
\label{fodo60}
\end{figure}

\begin{figure}[h!btp]
\setlength{\unitlength}{1mm}
\centering
\includegraphics*[width=80mm]{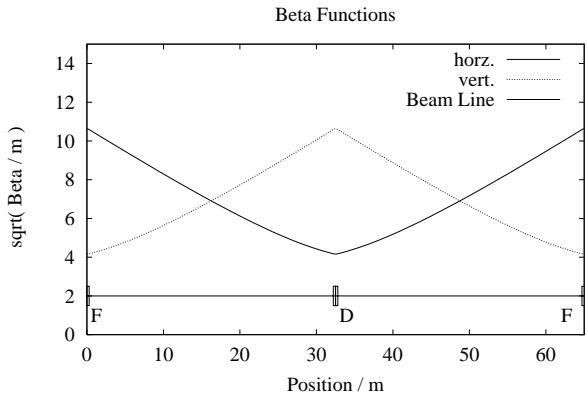}
\vspace*{-5mm}
\caption{The square root of the horizontal (solid line) and vertical
(dashed line) beta-function of the FEL beam. }
\label{fodo90}
\end{figure}

The beta-functions of a standard (i.e. length $L_{cell} = 65 \,{\rm m}$)
FODO cell with 60$^{\circ}$ phase advance is shown in Fig.~\ref{fodo60}.
The quadrupole strength is $ k = 0.0516 /{\rm m^2}$ for a quadrupole length
of $0.6$~m. The focusing effect of the cavities can be neglected for
a high energy beam. The minimal and maximal beta-functions are
$\check{\beta} = 37.58 \, {\rm m}$ and $\hat{\beta}= 112.41 \,
 {\rm m}$ for the HEP beam.
The focussing effect of the quadrupole is a factor $1.47$ stronger for
the FEL-beam resulting in beta-functions
 $\check{\beta} = 17.32 \, {\rm m}$ and $\hat{\beta}= 112.12 \, {\rm m}$
which are shown in Fig.~\ref{fodo90}.

\begin{figure}[h!btp]
\setlength{\unitlength}{1mm}
\centering
\includegraphics*[width=80mm]{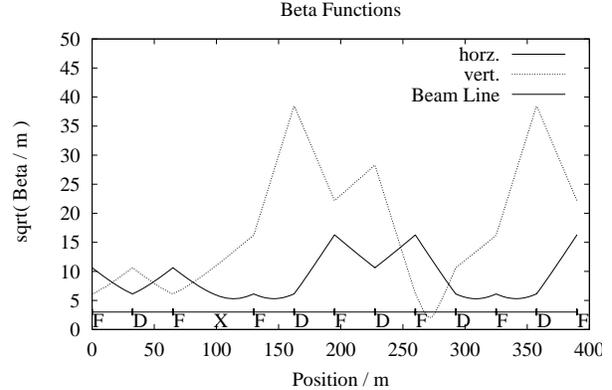}
\vspace*{-5mm}
\caption{Lattice with a missing quad.
The square root of the horizontal (solid line) and vertical
(dashed line) beta-function of the HEP beam are shown.}
\label{fodomis}
\end{figure}

\begin{figure}[h!btp]
\setlength{\unitlength}{1mm}
\centering
\includegraphics*[width=80mm]{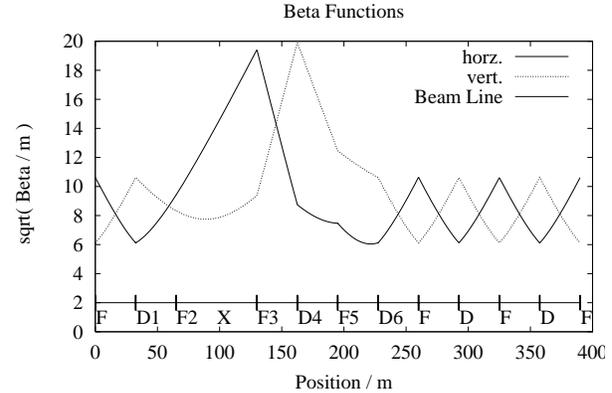}
\caption{Re-matched beta-functions with a missing quad.
The square root of the horizontal (solid line) and vertical
(dashed line) beta-function of the HEP beam are shown.}
\label{fodomat}
\end{figure}

Now it is assumed that one power supply failed causing a significant
betatron-mismatch (see Fig.~\ref{fodomis}). Using the computer
code COMFORT \cite{Comfort} it is possible to rematch the
beta-function such that two FODO cells down-stream from the
missing quadrupole the original periodic solution is recovered.
Actually there exist several solutions which rematch the lattice. The
solution presented in Fig.~\ref{fodomat} fulfills the
additional constraints that the strength of all quadrupoles does
not exceed the design values of the standard 60$^{\circ}$ lattice
and that the polarity is unchanged, i.e. an F(D)-quad remains an F(D)-quad.
The relative quadrupole strengths of the six matching quadrupoles
($k_n/k$, $n=1,\dots,6$) are summarized in table \ref{results}.

\begin{table}[htb]
\centerline{
\parbox[t]{3cm}{
\begin{tabular} {|c|c|}
        \hline
%
Quad       & $k_n/k$    \\
\hline
{\bf D 1}  & $ 0.7694 $ \\
{\bf F 2}  & $ 0.0    $ \\
{\bf F 3}  & $ 0.8283$ \\
\hline
\end{tabular}
} \ \hspace*{0.5cm}  \
\parbox[t]{3cm}{
\begin{tabular} {|c|c|}
        \hline
%
Quad       & $k_n/k$    \\
\hline
{\bf D 4}  & $ 0.9118 $ \\
{\bf F 5}  & $ 0.4056 $ \\
{\bf D 6}  & $ 0.3616$ \\
\hline
\end{tabular}
}}
\vspace{-1mm}
\caption{\label{results} Relative quadrupole strength of the 6
matching quadrupoles. All quadrupole strengths are normalized to the
quadrupole of a regular FODO-cell.}
\end{table}

While the solution presented in Fig.~\ref{fodomat} is perfect
for the HEP beam, the FEL beam is still mismatched. It is of
course possible to find another solution for the FEL beam but
that solution would not fulfill the matching conditions
for the HEP beam. In the next section a solution is shown which
matched the lattice for the HEP and FEL beam simultaneously.


\section{Simultaneous Lattice Matching  for the HEP and FEL Beam }

Simultaneous matching of the lattice for the HEP and FEL beams
requires a solution to the following four equations:
\begin{eqnarray}
 T(k_1, \dots, k_n) \, \left( \begin{array}{c}
             \beta_{HEP,j} \\ \alpha_{HEP,j} \\ \gamma_{HEP,j}
                       \end{array} \right)
  & = &
 \left( \begin{array}{c}
             \beta_{HEP,j} \\ \alpha_{HEP,j} \\ \gamma_{HEP,j}
                       \end{array} \right)
                                                  \nonumber \\
         & &   \hspace*{16mm} j=x,y             \nonumber \\
 T(\lambda \,k_1, \dots, \lambda  k_n) \, \left( \begin{array}{c}
             \beta_{FEL,j} \\ \alpha_{FEL,j} \\ \gamma_{FEL,j}
                       \end{array} \right)
  & = &
 \left( \begin{array}{c}
             \beta_{FEL,j} \\ \alpha_{FEL,j} \\ \gamma_{FEL,j}
                       \end{array} \right),
\nonumber
\end{eqnarray}
where $T$ is the $3 \times 3$ lattice transfer matrix depending
on the quadrupole strengths $k_i$ ($i=1 \dots n$), where
$\beta_{HEP,j}$, $\alpha_{HEP,j}$, $\gamma_{HEP,j}$,   
$\beta_{FEL,j}$, $\alpha_{FEL,j}$, $\gamma_{FEL,j}$
($j=x,y$) are the horizontal and vertical twiss parameter of
the HEP and FEL beam corresponding to the periodic
solutions of a standard cell (see Fig.~\ref{fodo60} and
Fig.~\ref{fodo90}), and $\lambda=1.47$ is the ratio of the beam
momenta of the HEP and FEL beam.

\begin{figure}[h!btp]
\setlength{\unitlength}{1mm}
\centering
\includegraphics*[width=80mm]{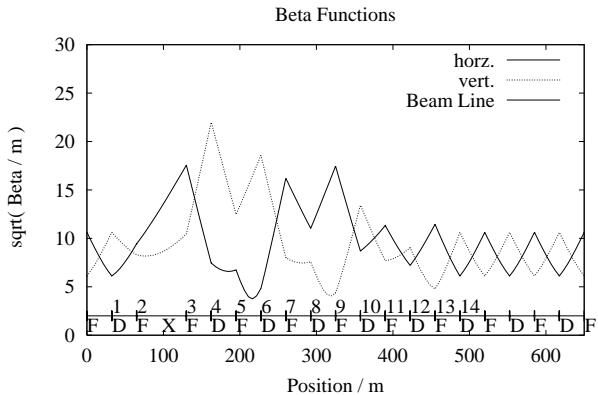}
\vspace*{-5mm}
\caption{Simultaneous beam matching with a missing quad.
The square root of the horizontal (solid line) and vertical
(dashed line) beta-function of the HEP beam is shown. }
\label{HEPmatch}
\end{figure}

\begin{figure}[h!btp]
\setlength{\unitlength}{1mm}
\centering
\includegraphics*[width=80mm]{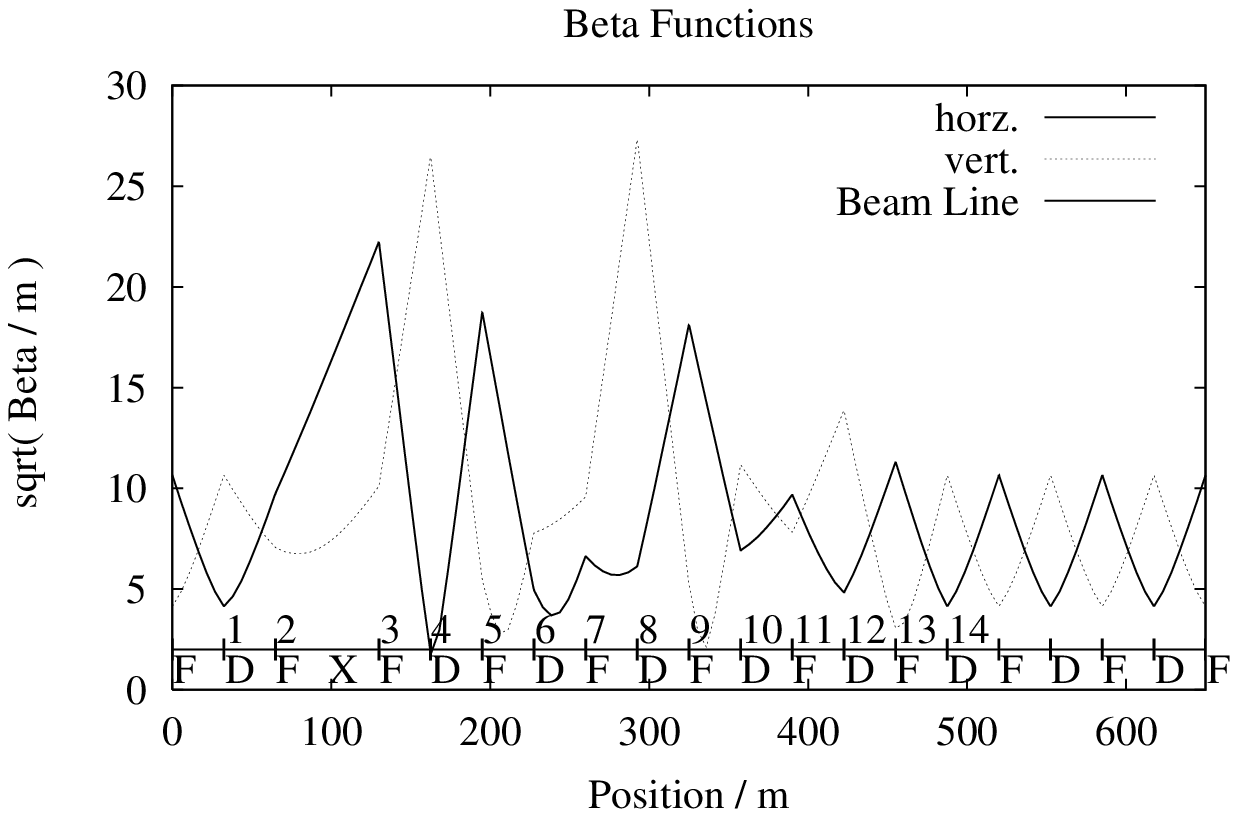}
\vspace*{-5mm}
\caption{Simultaneous beam matching with a missing quad.
The square root of the horizontal (solid line) and vertical
(dashed line) beta-function of the FEL beam is shown. }
\label{FELmatch}
\end{figure}

Commonly used beam optics computer codes like COMFORT cannot
match the beta-functions for two beams of different momenta
simultaneously. Using the code Mathcad \cite{Mathcad} it is
possible to solve the above equations numerically using $n=14$
quadrupoles. The algorithm is based on a quasi-Newton
method \cite{Mathcad}. The result of the computations is shown in
Fig.~\ref{HEPmatch} for the HEP beam and in Fig.~\ref{FELmatch} 
for the FEL beam. Please note that the magnetic gradient
of all quads is the same in both cases only the beam momenta
differ. The relative strengths of the 14 matching quadrupoles are
given in table \ref{HEPFELresults}. One quadrupole is 22~\%
stronger than a quadrupole of a regular FODO cell, which is
not critical since the simultaneous matching solution is only
required for beam energies up-to 50~GeV.

\begin{table}[htb]
\centerline{
\parbox[t]{3cm}{
\begin{tabular} {|c|c|}
        \hline
%
Quad       & $k_n/k$    \\
\hline
{\bf D 1}  & $ 0.1075 $ \\
{\bf F 2}  & $ 0.7727 $ \\
{\bf F 3}  & $ 0.8390 $ \\
{\bf D 4}  & $ 0.9646 $ \\
{\bf F 5}  & $ 1.2198 $ \\
{\bf D 6}  & $ 0.9129 $ \\
{\bf F 7}  & $ 1.0559 $ \\
\hline
\end{tabular}
} \ \hspace*{0.5cm}  \
\parbox[t]{3cm}{
\begin{tabular} {|c|c|}
        \hline
%
Quad       & $k_n/k$    \\
\hline
{\bf D 8}  & $ 0.9962 $ \\
{\bf F 9}  & $ 0.8883 $ \\
{\bf D 10} & $ 1.1831 $ \\
{\bf F 11} & $ 0.6901 $ \\
{\bf D 12} & $ 0.8955 $ \\
{\bf F 13} & $ 0.9455 $ \\
{\bf D 14} & $ 1.1530 $ \\
\hline
\end{tabular}
}}
\vspace{-0.1cm}
\caption{\label{HEPFELresults} Relative quadrupole strengths of the 14
matching quadrupoles. All quadrupole strengths are normalized to the
quadrupole of a regular FODO-cell.}
\end{table}

\section{Conclusion}

It has been demonstrated that there exist matched beam
optics with a missing quadrupole for the HEP beam and
even simultaneously for the HEP and FEL beam.
Therefore  a power supply failure will not cause a
significant down-time of the linear accelerator since
the operation can continue with a re-matched lattice.
\begin{center}
{\bf Acknowledgment }
\end{center}
I would like to thank R.~Brinkmann for discussions
and suggestion for matching of the 60$^{\circ}$-lattice.
Thanks go also to S.G.~Wipf for carefully reading the manuscript.



\end{document}